\documentclass[11pt]{article}
\pdfoutput=1  
\usepackage{graphicx,color}
\usepackage{appendix}
\usepackage{latexsym,amsmath,amssymb,graphicx,booktabs}
\usepackage{epsfig,latexsym,cite}
\usepackage{hyperref}
\numberwithin{equation}{section}

\definecolor{MyBlue}{rgb}{0.15,0.15,0.70}

\hypersetup{
colorlinks=true,
citecolor=MyBlue,
linkcolor=MyBlue,
urlcolor=MyBlue
}

\input epsf
\usepackage{amssymb}
\usepackage{amsmath}
\usepackage{amsfonts}
\usepackage{upgreek}
\usepackage{latexsym}

\newcommand{\iBox}{\Box^{-1}}

\renewcommand\({\left(}
\renewcommand\){\right)}
\renewcommand\[{\left[}
\renewcommand\]{\right]}

\newcommand{\ra}{\rightarrow}

\def\lsim{\raise 0.4ex\hbox{$<$}\kern -0.8em\lower 0.62
ex\hbox{$\sim$}}

\def\gsim{\raise 0.4ex\hbox{$>$}\kern -0.7em\lower 0.62
ex\hbox{$\sim$}}

\def\lbar{{\hbox{$\lambda$}\kern -0.7em\raise 0.6ex
\hbox{$-$}}}

\newcommand\eq[1]{eq.~(\ref{#1})}
\newcommand\eqs[2]{eqs.~(\ref{#1}) and (\ref{#2})}
\newcommand\Eq[1]{Equation~(\ref{#1})}

\newcommand\eqsss[4]{eqs.~(\ref{#1}), (\ref{#2}), (\ref{#3})
and (\ref{#4})}

\newcommand\eqst[2]{eqs.~(\ref{#1})--(\ref{#2})}

\newcommand\pa{\partial}
\newcommand\p{\partial}

\newcommand\ee{\end{equation}}
\newcommand\be{\begin{equation}}
\def\bea{\begin{array}}
\def\eea{\end{array}}\def\ea{\end{array}}
\newcommand\ees{\end{eqnarray}}
\newcommand\bees{\begin{eqnarray}}
\def\nn{\nonumber}

\def\a{\alpha}

\def\d{\delta}

\def\eps{\epsilon}

\def\dslash{\hspace{-1mm}\not{\hbox{\kern-2pt $\partial$}}}
\def\Dslash{\not{\hbox{\kern-4pt $D$}}}
\def\pslash{\not{\hbox{\kern-2.1pt $p$}}}
\def\kslash{\not{\hbox{\kern-2.3pt $k$}}}
\def\qslash{\not{\hbox{\kern-2.3pt $q$}}}

\def\p1{{\bf p}_1}
\def\p2{{\bf p}_2}
\def\k1{{\bf k}_1}
\def\k2{{\bf k}_2}

\newcommand{\emn}{\eta_{\mu\nu}}

\newcommand{\eMN}{\eta^{\mu\nu}}

\newcommand{\gmn}{g_{\mu\nu}}

\newcommand{\gbmn}{\bar{g}_{\mu\nu}}

\newcommand{\fbmn}{\bar{f}_{\mu\nu}}

\newcommand{\hmn}{h_{\mu\nu}}
\newcommand{\hrs}{h_{\rho\sigma}}
\newcommand{\hmr}{h_{\mu\rho}}

\newcommand{\hnr}{h_{\nu\rho}}

\newcommand{\fmn}{f_{\mu\nu}}
\newcommand{\lmn}{l_{\mu\nu}}

\newcommand{\bhmn}{\bar{h}_{\mu\nu}}

\newcommand{\pam}{\pa_{\mu}}

\newcommand{\pan}{\pa_{\nu}}
\newcommand{\parho}{\pa_{\rho}}

\newcommand{\paM}{\pa^{\mu}}
\newcommand{\paN}{\pa^{\nu}}
\newcommand{\paR}{\pa^{\rho}}
\newcommand{\paS}{\pa^{\sigma}}

\newcommand{\Rmn}{R_{\mu\nu}}

\newcommand{\RMN}{R^{\mu\nu}}

\newcommand{\Rmnrs}{R_{\mu\nu\rho\sigma}}

\newcommand{\Tmn}{T_{\mu\nu}}

\newcommand{\TMN}{T^{\mu\nu}}

\newcommand{\dddM}{\kern 0.2em \raise 1.9ex\hbox{$...$}\kern -1.0em \hbox{$M$}}
\newcommand{\dddQ}{\kern 0.2em \raise 1.9ex\hbox{$...$}\kern -1.0em \hbox{$Q$}}
\newcommand{\dddI}{\kern 0.2em \raise 1.9ex\hbox{$...$}\kern -1.0em\hbox{$I$}}
\newcommand{\dddJ}{\kern 0.2em \raise 1.9ex\hbox{$...$}\kern-1.0em
\hbox{$J$}}
\newcommand{\dddcalJ}{\kern 0.2em \raise 1.9ex\hbox{$...$}\kern-1.0em
\hbox{${\cal J}$}}

\newcommand{\dddO}{\kern 0.2em \raise 1.9ex\hbox{$...$}\kern -1.0em
\hbox{${\cal O}$}}
\def\dddz{\raise 1.5ex\hbox{$...$}\kern -0.8em \hbox{$z$}}
\def\dddd{\raise 1.8ex\hbox{$...$}\kern -0.8em \hbox{$d$}}
\def\dddbd{\raise 1.8ex\hbox{$...$}\kern -0.8em \hbox{${\bf d}$}}
\def\ddbd{\raise 1.8ex\hbox{$..$}\kern -0.8em \hbox{${\bf d}$}}
\def\dddx{\raise 1.6ex\hbox{$...$}\kern -0.8em \hbox{$x$}}

\begin{document}

\begin{titlepage}

\vspace*{2cm}

\centerline{\Large \bf Non-local formulation of   ghost-free bigravity theory}

\vskip 0.4cm
\vskip 0.7cm
\centerline{\large    Giulia Cusin$^a$, Jacopo Fumagalli$^{a,b,c}$ and Michele Maggiore$^a$ }
\vskip 0.3cm
\centerline{\em $^a$D\'epartement de Physique Th\'eorique and Center for Astroparticle Physics,}  
\centerline{\em Universit\'e de Gen\`eve, 24 quai Ansermet, CH--1211 Gen\`eve 4, Switzerland}
\centerline{\em $^b$Dipartimento di Fisica,
Universit\`a degli Studi di Pavia,
via Bassi 6, 27100 Pavia, Italy}
\centerline{\em $^c$INFN Sezione di Pavia,
via Bassi 6, 27100 Pavia, Italy}

\vskip 1.9cm

\begin{abstract}

We study the ghost-free bimetric theory of Hassan and Rosen, with parameters $\beta_i$  such that a flat Minkowski solution exists for both metrics. We show that, expanding around this solution and eliminating one of the two metrics with its own equation of motion, the remaining metric is governed by 
the Einstein-Hilbert action plus a non-local term proportional to $W_{\mu\nu\rho\sigma} (\Box-m^2)^{-1}W^{\mu\nu\rho\sigma}$, where $W_{\mu\nu\rho\sigma}$ is the Weyl tensor. The result is valid 
to quadratic order in the metric perturbation and to all orders in the derivative expansion.
This example shows, in a simple setting, how such non-local extensions of GR can emerge from an underlying consistent theory, at the purely classical level.

\end{abstract}

\end{titlepage}

\newpage

\section{Introduction}

The study of infrared modifications of General Relativity (GR) is motivated both by its intrinsic conceptual interest and by the aim of explaining the observed accelerated expansion of the Universe. A natural way of modifying the theory in the infrared is to add a mass term. In this direction, significant progresses have been made in recent years with the construction of a ghost-free theory of massive gravity, the dRGT theory~\cite{deRham:2010ik,deRham:2010kj} (see also \cite{deRham:2010gu,deRham:2011rn,deRham:2011qq,Hassan:2011hr,Hassan:2011vm,Hassan:2011tf,Hassan:2011ea,Hassan:2012qv,Comelli:2012vz,Jaccard:2012ut,Comelli:2012db,Comelli:2013txa,Guarato:2013gba}, and\cite{Hinterbichler:2011tt,deRham:2014zqa} for   reviews). Such a theory involves, beside the dynamical metric $\gmn$, a non-dynamical reference metric $\fmn$ which is needed to construct a mass term. A natural subsequent step is to promote $\fmn$ to a dynamical field. This leads to  bimetric theories.  Ghost-free massive gravity has been generalized to a ghost-free bimetric theory by Hassan and Rosen \cite{Hassan:2011zd}. Conceptual aspects  of bigravity have been  investigated  in  \cite{Hassan:2012gz,Hassan:2012wr,Hassan:2013pca},
and its cosmological consequences have been studied  e.g. in \cite{vonStrauss:2011mq,Tamanini:2013xia,Fasiello:2013woa,Akrami:2013ffa,Konnig:2013gxa,Konnig:2014dna,Comelli:2014bqa,Solomon:2014dua,DeFelice:2014nja}.
The Hassan-Rosen bimetric theory is defined by the action
\be\label{eq:1}
S=\int  d^4x\, M_{g}^{2}\sqrt{-g}R(g)+\int d^4x\,  M_{f}^{2}\sqrt{-f}R(f)-M_{f}^{2}m^{2}\int d^4x\, \sqrt{-g}\sum_{n=0}^{4}\beta_{n}e_{n}(\mathbb{X})\, ,
\ee
where  $\beta_{i}$ are general real coefficients,  $m$ is a parameter with the dimension of mass, 
$\mathbb{X_{\mu}^{\nu}=\mathrm{(}\sqrt{\mathrm{g^{-1}f}}})_{\mu}^{\nu}$
and the  $e_{i}(\mathbb{X})$ polynomials are given by
\be
\begin{array}{c}
\begin{array}{c}
e_{0}=I\;\; e_{1}=[\mathbb{X}]\;\; e_{2}=\frac{1}{2}([\mathbb{X}]^{2}-[\mathbb{X}^{2}])\;\; e_{3}=\frac{1}{6}([\mathbb{X}]^{3}-3[\mathbb{X}][\mathbb{X}^{2}]+2[\mathcal{\mathbb{X}}^{3}]\\
\\
e_{4}=\frac{1}{24}([\mathbb{X}]^{4}-6[\mathbb{X}]^{2}[\mathbb{X}^{2}]+8[\mathbb{X}][\mathbb{X}^{3}]+3[\mathbb{X}^{2}]^{2}-6[\mathbb{X}^{4}])\, ,
\end{array}\end{array}
\ee
where the bracket denotes the trace of the matrix and, for simplicity, we have restricted ourselves to $D=4$ space-time dimensions. 

The purpose of this paper is to show how this theory can be recast into a non-local form involving only one metric. Working  up to terms quadratic in the curvature and choosing the parameters $\beta_i$ such that the theory admits a background solution $\gbmn=\fbmn=\emn$, we will find that the action (\ref{eq:1}) is classically equivalent to the action
\be\label{Sprime}
S'=M^2_{\rm pl}\int d^4x\sqrt{-g}\, R(g)\,
-\frac{M_{f}^{2}}{2}\int d^4x\,\sqrt{-g}\,
W_{\mu\nu\rho\sigma} \frac{1}{\Box-m^{2}}W^{\mu\nu\rho\sigma}+{\cal O}(\Rmnrs^3)\, ,
\ee
where $W^{\mu\nu\rho\sigma}$ is the Weyl tensor constructed with the metric $\gmn$, and $M^2_{\rm pl}=M_g^2+M_f^2$.
 In order to get this result, we will integrate out $\fmn$ by using its own equations of motion, linearized over Minkowski, and we will then covariantize the result.\footnote{In the following, in fully covariant expressions it is understood that $\Box$ is the d'Alembertian computed with respect to the full metric $\gmn$, while in linearized expression it is understood that
$\Box=\eMN\pam\pan$ is the flat-space d'Alembertian, and similarly for its inverse $\iBox$.} Our analysis will complement the study performed by Hassan, Schmidt-May and von~Strauss~\cite{Hassan:2013pca}, where $\fmn$ is rather eliminated using the equation of motion of $\gmn$; we will comment below on the relation between the two approaches.

This result  reveals an interesting relation between  bigravity and Stelle's higher derivative gravity. 
The term $W_{\mu\nu\rho\sigma} (\Box-m^{2})^{-1}W^{\mu\nu\rho\sigma}$ can be seen as a UV completion of a term $-(1/m^2)W_{\mu\nu\rho\sigma} W^{\mu\nu\rho\sigma}$. 
In the infrared limit  $(\Box-m^{2})^{-1}\simeq -1/m^2$ and, neglecting also cubic and higher order terms,
\eq{Sprime} reduces to
\be\label{Stelle}
S_{{\rm Stelle}}=M^2_{\rm pl}\int d^4x\sqrt{-g}\, R(g)\, 
+c_W
\int d^4x\,\sqrt{-g}\,
W_{\mu\nu\rho\sigma} W^{\mu\nu\rho\sigma}\, ,
\ee
(where $c_W= M_f^2/(2m^2)$), which is
the action of Stelle's theory~\cite{Stelle:1977ry,Stelle:1976gc}. 
Stelle's theory has 7 propagating degrees of freedom, organized into a massless spin-2 graviton and a massive ghost-like spin-2 state. The original bigravity theory also has a massless and a massive graviton, but is  ghost-free. Therefore, this construction provides an explicit example of how to embed Stelle's higher-derivative gravity into a ghost-free theory (as already discussed from a different point of view  in~\cite{Hassan:2013pca}). 
The  non-local expression (\ref{Sprime}) is also  useful to investigate the relation, and the differences, between this non-local formulation of bigravity, and  non-local modifications of General Relativity such as those that have been discussed   in \cite{Jaccard:2013gla,Maggiore:2013mea,Maggiore:2014sia,Foffa:2013sma,Foffa:2013vma,Kehagias:2014sda,Nesseris:2014mea,Dirian:2014ara,Conroy:2014eja}.

The paper is organized as follows. In sect.~\ref{sect:elim}  we express the fluctuations of the metric $\fmn$ in terms of that of $\hmn$. The non-local action is computed in sect.~\ref{sect:NLaction}. We conclude with a discussion of our results in sect.~\ref{sect:disc}. 
In app.~\ref{sect:compare} we compare our results with that of ref.~\cite{Hassan:2013pca} and in
app.~\ref{app:B} we extend the computation to the interaction with matter. 
We use the signature $(-,+,+,+)$ and units $\hbar=c=1$.

\section{Elimination of the second metric}\label{sect:elim}

\subsection{Computation of $\lmn$}

The  equations of motion derived from \eq{eq:1} are
\bees
\frac{M_{g}^{2}}{M_{f}^{2}}G_{\mu\nu}(g)+m^{2}\sum_{n=0}^{3}(-1)^{n}\beta_{n}g_{\mu\lambda}\mathbb{Y}_{(n)\nu}^{\lambda}(\mathbb{X})&=&0\, ,\label{dSdg}\\
G_{\mu\nu}(f)+m^{2}\sum_{n=0}^{3}(-1)^{n}\beta_{4-n}f_{\mu\lambda}\mathbb{Y}_{(n)\nu}^{\lambda}(\mathbb{X}^{-1})&=&0\, ,\label{dSdf}
\ees
where  $\mathbb{Y}_{(n)}(\mathbb{X})=\sum_{r=0}^n(-1)^r\mathbb{X}^{n-r}e_r(\mathbb{X})$, and we neglect for the moment matter sources (the extension to matter sources is performed in
app.~\ref{app:B}). In order to obtain an effective action involving only the metric $\gmn$, we eliminate $\fmn$ by using its own equation of motion. This involves the inversion of a differential operator, which  in practice can only be done by expanding around a simple background, such as Minkowski. The result can then be covariantized which, as long as one truncates the theory to quadratic order, can be done uniquely. 
Thus, 
in order to simplify the problem, we choose the coefficients $\beta_n$ in \eq{eq:1} in such a way that there exists a solution
of the equations of motion with $\gbmn=\fbmn=\emn$. This can be obtained for instance setting~\cite{Hassan:2012wr}
\begin{equation}
\beta_{0}=\beta_{4}+2\beta_{3}-2\beta_{1};
\;\;\;\beta_{2}=-\frac{\beta_{1}}{3}-\frac{\beta_{4}}{3}-\beta_{3};
\hspace{4mm}\;\;(\beta_{1},\beta_{3},\beta_{4})\in\mathbb{R}\label{eq:4}
\end{equation}
Imposing that only one of the three remaining free parameters i.e.
$\beta_{1}$ is different from zero, \eq{eq:4} implies 
$\beta_{0}=-2\beta_{1}$ and $\beta_{2}=-\beta_{1}/3$. In the following we adopt for definiteness this choice and, for later convenience, we set  $\beta_1=3$ (in any case, different choices of the $\beta_i$, satisfying
\eq{eq:4},  can be reabsorbed in the definition of $m^2$). Then the potential term in \eq{eq:1} becomes
\be
\sum_{n=0}^{3}\beta_{n}e_{n}(\mathbb{X})=  -6e_{0}(\mathbb{X})+3e_{1}(\mathbb{X})
-e_{2}(\mathbb{X})\, ,
\ee
and the equations for $f_{\mu\nu}$ 
and  $g_{\mu\nu}$  become 
\bees
\frac{M_{g}^{2}}{M_{f}^{2}}G_{\mu\nu}(g)-m^{2}\[6g_{\mu\lambda}\mathbb{Y}_{0\nu}^{\lambda}(\mathbb{X})+3g_{\mu\lambda}\mathbb{Y}_{1\nu}^{\lambda}(\mathbb{X})+g_{\mu\lambda}\mathbb{Y}_{2\nu}^{\lambda}(\mathbb{X})\]&=&0\, ,\label{eq:5a}\\
G_{\mu\nu}(f)-m^{2}\[ 3 f_{\mu\lambda}\mathbb{Y}_{3\nu}^{\lambda}(\mathbb{X}^{-1})+f_{\mu\lambda}\mathbb{Y}_{2\nu}^{\lambda}(\mathbb{X}^{-1})\]&=&0\, .\label{eq:5b}
\ees
We now expand
$g_{\mu\nu}=\eta_{\mu\nu}+h_{\mu\nu}$, $f_{\mu\nu}=\eta_{\mu\nu}+l_{\mu\nu}$. Then
\eqs{eq:5a}{eq:5b} give
\bees
\frac{M_{g}^{2}}{M_{f}^{2}}\mathcal{E}_{\mu\nu\rho\sigma}h^{\rho\sigma}+m^{2}(h_{\mu\nu}-\eta_{\mu\nu}h)&=&m^{2}(l_{\mu\nu}-\eta_{\mu\nu}l)\, ,\label{eq:6a}\\
\mathcal{E}_{\mu\nu\rho\sigma}l^{\rho\sigma}+m^{2}(l_{\mu\nu}-\eta_{\mu\nu}l)&=&m^{2}(h_{\mu\nu}-\eta_{\mu\nu}h)\, ,\label{eq:6b}
\ees
where for the Lichnerowicz operator we use the convention
\be
\mathcal{E}_{\mu\nu\rho\sigma}h^{\rho\sigma}=-\Box h_{\mu\nu}+\eta_{\mu\nu}\Box h-\partial_{\mu}\partial_{\nu}h-\eta_{\mu\nu}\partial_{\rho}\partial_{\sigma}h^{\rho\sigma}+\partial^{\rho}\partial_{\nu}h_{\rho\mu}+\partial^{\rho}\partial_{\mu}h_{\rho\nu}\, .
\ee
It is also convenient to define the tensor
\be
S_{\mu\nu\rho\sigma}=\frac{1}{2}\( \eta_{\mu\rho}\eta_{\nu\sigma}+\eta_{\mu\sigma}\eta_{\nu\rho}\)
-\eta_{\mu\nu}\eta_{\rho\sigma}\, .\label{eq:tensore S}
\ee
\Eq{eq:6b} can then be rewritten as
\be\label{eq:7}
(\mathcal{E}_{\mu\nu\rho\sigma}+m^{2}S_{\mu\nu\rho\sigma})l^{\rho\sigma}=m^{2}S_{\mu\nu\rho\sigma}h^{\rho\sigma}\, .
\ee
The operator acting on $l$
is precisely the Fierz-Pauli operator. We know that for $m\neq 0$ it is invertible and the inverse is
\begin{equation}\label{Q}
Q_{\mu\nu\rho\sigma}=-\frac{1}{\Box-m^{2}}\[
\frac{1}{2}\( \Pi_{\mu\rho}\Pi_{\nu\sigma}+\Pi_{\mu\sigma}\Pi_{\nu\rho}\)
-\frac{1}{3}\Pi_{\mu\nu}\Pi_{\rho\sigma}\]\, ,
\end{equation}
where $\Pi_{\mu\nu}=\emn- m^{-2}\pam\pan$. Then
\bees
l^{\mu\nu}&=&m^{2}Q^{\mu\nu\alpha\beta}S_{\alpha\beta\rho\sigma}h^{\rho\sigma}\nn\\
&=&\frac{1}{\Box -m^{2}}
\( \pa^{\mu}\pa_{\alpha}h^{\nu\alpha}+\pa^{\nu}\pa_{\alpha}h^{\mu\alpha}-\pa^{\mu}\pa^{\nu}h\)
-\frac{m^{2}}{\Box-m^{2}}h^{\mu\nu}\nn\\
&&+\frac{1}{3}\frac{1}{\Box-m^{2}}\(\eta^{\mu\nu}+\frac{2\pa^{\mu}\pa^{\nu}}{m^{2}}\)
(\Box h-\pa_{\alpha}\pa_{\beta}h^{\alpha\beta})\, .\label{lmnhmn1}
\ees
This expression can be rewritten in terms of the linearized Ricci tensor $\mathcal{R}_{\mu\nu}$ and of the linearized Ricci scalar $\mathcal{R}$ (we use calligraphic letters to denote quantities linearized over Minkowski),
which are given by
\bees
\mathcal{R}_{\mu\nu}&=&\frac{1}{2}\( \pa^{\a}\pam h_{\nu\a}+\pa^{\a}\pan h_{\mu\a}
-\Box\hmn-\pam\pan h\)\, ,\label{linRmn1}\\
\mathcal{R}&=&\paM\paN\hmn-\Box h\, .\label{linRmn2}
\ees
Then \eq{lmnhmn1} becomes
\be\label{eq:10}
l_{\mu\nu}=h_{\mu\nu}+\frac{1}{\Box-m^{2}}\[2\mathcal{R}_{\mu\nu}-\frac{1}{3}\(\eta_{\mu\nu}+2\frac{\partial_{\mu}\partial_{\nu}}{m^{2}}\)\mathcal{R}\]\, .
\ee
Taking the trace we get
\begin{equation}
l=h-\frac{2\mathcal{R}}{3m^{2}}\, ,\label{eq:11}
\end{equation}
so the trace $l$ is a local function of $\hmn$.  Plugging  these two expressions in \eq{eq:6a} we get
a non local equation for $\hmn$,
\begin{equation}
\frac{M_{g}^{2}}{M_{f}^{2}}{\mathcal{E}_{\mu\nu}}^{\rho\sigma}h_{\rho\sigma}-\frac{2}{3}\eta_{\mu\nu}\mathcal{R}-\frac{m^{2}}{\Box-m^{2}}\[2\mathcal{R}_{\mu\nu}-\frac{1}{3}\(\eta_{\mu\nu}+2\frac{\partial_{\mu}\partial_{\nu}}{m^{2}}\)\mathcal{R}\]=0\, .\label{eq:12}
\end{equation}
It is straightforward to check that the divergence of the left-hand side vanishes identically, as it should. Therefore, when $\hmn$ is coupled to the matter energy-momentum tensor $\Tmn$, energy-momentum  conservation, $\paM\Tmn=0$, is automatically assured.

\subsection{Helicity decomposition of the metric perturbations}

It is instructive to repeat the above computation by first decomposing the metric perturbations $\hmn$ and $\lmn$ into their scalar, vector and tensor components, 
\bees
h_{\mu\nu}&=&h_{\mu\nu}^{TT}+\frac{1}{2}(\partial_{\mu}\epsilon_{\nu}^{T}+\partial_{\nu}\epsilon_{\mu}^{T})+\partial_{\mu}\partial_{\nu}\alpha+\frac{1}{3}\eta_{\mu\nu}s\, ,\label{eq:nonlocal1}\\
l_{\mu\nu}&=&l_{\mu\nu}^{TT}+\frac{1}{2}(\partial_{\mu}l_{\nu}^{T}+\partial_{\nu}l_{\mu}^{T})+\partial_{\mu}\partial_{\nu}\beta+\frac{1}{3}\eta_{\mu\nu}u\, ,\label{eq:nonlocal2}
\ees
where  $h_{\mu\nu}^{TT}$
is the transverse-traceless part, $\partial_{\mu}h_{\mu\nu}^{TT}=0$, $\eta^{\mu\nu}h_{\mu\nu}^{TT}=0$,
$\epsilon_{\mu}^{T}$ is a transverse vector, $\partial^{\mu}\epsilon_{\mu}^{T}=0$, and $\alpha$ and $s$ are scalar under rotation (and similarly for the decomposition of $\lmn$).
We also define $v_{\mu\nu}=h_{\mu\nu}-l_{\mu\nu}$ and we decompose it as
\begin{equation}
v_{\mu\nu}=v_{\mu\nu}^{TT}+\frac{1}{2}(\partial_{\mu}v_{\nu}^{T}+\partial_{\nu}v_{\mu}^{T})+\partial_{\mu}\partial_{\nu}\gamma+\frac{1}{3}\eta_{\mu\nu}c
\end{equation}
so, of course, $v_{\mu\nu}^{TT}=h_{\mu\nu}^{TT}-l_{\mu\nu}^{TT}$,  $v_{\mu}^{T}=\epsilon_{\mu}^{T}-l_{\mu}^{T}$, $\gamma=\alpha-\beta$ and $c=s-u$. In term of these variables the quadratic Einstein-Hilbert actions take the  form (see e.g. \cite{Hassan:2011tf} or app.~B of \cite{Jaccard:2013gla})

\begin{equation}
S^{(2)}_{EH_{1}}+S^{(2)}_{EH_{2}}=\frac{1}{4}\int d^{4}x\:\: M_{g}^{2}\(h_{\mu\nu}^{TT}\Box h_{\mu\nu}^{TT}-\frac{2}{3}s\Box s\)
+M_{f}^{2}\(l_{\mu\nu}^{TT}\Box l_{\mu\nu}^{TT}-\frac{2}{3}u\Box u\)\, ,
\end{equation}
while, after some integrations by part, the interaction term coming from the dRGT potential takes the form
\begin{equation}\label{Sint1}
S_{\rm int}=\frac{-M_{f}^{2}m^{2}}{4}\int d^4x\, \(v_{\mu\nu}^{TT}v^{TT\mu\nu}-\frac{1}{2}v_{\mu}^{T}\Box v^{T\mu}-2c\Box\gamma -\frac{4}{3}c^{2}\)\, .
\end{equation}
The corresponding equations of motion are
\bees
\Box v^{T\mu}&=&0\, ,\label{BoxvT}\\
\Box c&=&0\, ,\label{Boxc}\\
(\Box-m^2) l_{\mu\nu}^{TT}&=&-m^{2}h_{\mu\nu}^{TT}\, ,\\
\frac{2}{3}\Box u+m^{2}\(\Box\gamma+\frac{4}{3}c\)&=&0\, .
\ees
\Eq{BoxvT} implies $\Box  l_{\mu}^{T}=\Box \epsilon_{\mu}^{T}$. We solve it with the boundary condition that, when $ \epsilon_{\mu}^{T}=0$, we must have $l_{\mu}^{T}=0$. Then, $\Box v^{T\mu}=0$ implies 
$v^{T\mu}=0$.
Similarly, $\Box c=0$ implies $c=0$. Therefore we get 
\be\label{lepsus}
l_{\mu}^{T}=\epsilon_{\mu}^{T}\, ,\qquad u=s \, .
\ee
The other two equations give
\be\label{lTT1}
l_{\mu\nu}^{TT}=-\frac{m^{2}}{\Box-m^{2}}h_{\mu\nu}^{TT}
\ee
and 
\be\label{lTT2}
\beta=\alpha+\frac{2s}{3m^{2}}\, .
\ee
This decomposition allows us to appreciate that the non-locality in the relation between $\hmn$ and $\lmn$ only appears  in the tensor sector. The equivalence with the result found in \eq{eq:10} is easily proved inverting the decomposition (\ref{eq:nonlocal1}), which  gives (in $D=4$ space-time dimensions)~\cite{Jaccard:2013gla}
\bees
\alpha&=&-\frac{1}{3}\frac{1}{\Box}\(\eta^{\mu\nu}-\frac{4}{\Box}\partial^{\mu}\partial^{\nu}\)h_{\mu\nu}\label{eq:alpha ex}\, ,\\
s&=&\(\eta^{\mu\nu}-\frac{1}{\Box}\partial^{\mu}\partial^{\nu}\)h_{\mu\nu}\, , \label{eq:s=h+}\\
\eps_{\mu}^T&=&\frac{2}{\Box}\(\d_{\mu}^{\rho}-\frac{\pam\paR}{\Box}\)\paS\hrs\, ,\label{eqepsT}\\
\hmn^{\rm TT}&=&\hmn -\frac{1}{3}\(\emn -\frac{\pam\pan}{\Box}\)h
-\frac{1}{\Box}(\pam\paR\hnr+\pan\paR\hmr)\nn\\
&&+\frac{1}{3}\,\emn \frac{1}{\Box}\paR\paS\hrs 
+\frac{2}{3}\frac{1}{\Box^2}\pam\pan\paR\paS\hrs\, .\label{defhath}
\ees
Under linearized diffeomorphisms $\hmn\ra\hmn-(\pam\xi_{\nu}+\pan\xi_{\mu})$, decomposing 
$\xi_{\mu}=\xi_{\mu}^T+\pam\xi$, we have $\eps_{\mu}^{\rm T}\ra\eps_{\mu}^{\rm T}-2\xi_{\mu}^T$ and
$\a\ra \a-2\xi$, while $\hmn^{\rm TT}$ and $s$ are invariant.
Thus we can choose the gauge so that $\eps^T_{\mu}=\a=0$, and this leaves no residual gauge symmetry. 
Since $\hmn^{\rm TT}$ and $s$ are invariant, it is possible
to express them in terms of the linearized Ricci scalar and Ricci tensor (recall that, in linearized theory, the Riemann tensor is gauge-invariant rather than covariant). Indeed, \eqs{eq:s=h+}{defhath} can be rewritten as 
\bees
s&=&-\frac{1}{\Box}\mathcal{R}\label{eq:s=R/box}\, ,\\
h_{\mu\nu}^{TT}&=&\frac{2}{3}\frac{\partial_{\mu}\partial_{\nu}}{\Box^{2}}\mathcal{R}+\frac{1}{3}\frac{\eta_{\mu\nu}}{\Box}\mathcal{R}-\frac{2}{\Box}\mathcal{R}_{\mu\nu}\label{eq:hTT}\, .
\ees
Substituting \eqst{lepsus}{lTT2}  into 
\eq{eq:nonlocal2}, and expressing $\alpha$, $s$, $\eps_{\mu}^T$ and $\hmn^{\rm TT}$ in terms of $\hmn$ using
\eqsss{eq:alpha ex}{eqepsT}{eq:s=R/box}{eq:hTT} it is straightforward to show that one recovers \eq{eq:10}.

\section{Non-local action}\label{sect:NLaction}

We can now describe the dynamics entirely in term of $\hmn$. Note that, since $\lmn$ has been expressed in terms of $\hmn$ by using its own equation of motion, it is legitimate to substitute \eq{eq:10} (or, equivalently, \eqst{lepsus}{lTT2}) directly into the action. We find  convenient to work with the variables that appear in the helicity decomposition. 
The quadratic Einstein-Hilbert term of the second metric becomes
\be
S_{EH_{2}}=\frac{M_{f}^{2}}{4}\int d^4x\, \[ m^{4}h_{\mu\nu}^{TT}\frac{1}{(\Box-m^{2})}h_{\mu\nu}^{TT}+m^{6}h_{\mu\nu}^{TT}\frac{1}{(\Box-m^{2})^{2}}h_{\mu\nu}^{TT}-\frac{2}{3}s\Box s\] \, .
\ee
The term  $S_{\rm int}$, given in \eq{Sint1},  greatly simplifies thanks to \eq{lepsus}, and becomes
\be
S_{\rm int}=-\frac{M_{f}^{2}m^{2}}{4}\int d^4x\, \[ h_{\mu\nu}^{TT}h^{TT\mu\nu}+m^{4}h_{\mu\nu}^{TT}\frac{1}{(\Box-m^{2})^{2}}h^{TT\mu\nu}+2m^{2}h_{\mu\nu}^{TT}\frac{1}{\Box-m^{2}}h^{TT\mu\nu}\].
\ee
Since we have solved the equation for  $l_{\mu\nu}$
without the need of fixing the gauge, the resulting non-local action for $\hmn$ is 
invariant under linearized diffeomorphisms, and in fact it depends only on the invariant quantities
$h_{\mu\nu}^{TT}$ and  $s$. We can now use \eqs{eq:s=R/box}{eq:hTT} and, upon use of the linearized Bianchi identity $\paM \mathcal{R}_{\mu\nu}=(1/2)\pan \mathcal{R}$, we  get
\be
S^{(2)}_{EH_{2}}+S_{\rm int}=-M_{f}^{2}\int d^4x\, \[ \frac{1}{6}\mathcal{R}\frac{1}{\Box}\mathcal{R}+\mathcal{R}_{\mu\nu}\frac{m^{2}}{\Box(\Box-m^{2})}\mathcal{R}^{\mu\nu}-\tfrac{1}{3}\mathcal{R}\frac{m^{2}}{\Box(\Box-m^{2})}\mathcal{R}\] \, .
\ee
Observe that, since the term in square bracket is already ${\cal O}(h^2)$, at the quadratic order at which we are working we could simply replace $d^4x$ by $d^4x\sqrt{-g}$.
Using $\tfrac{m^{2}}{\Box(\Box-m^{2})}=-\tfrac{1}{\Box}+\tfrac{1}{\Box-m^{2}}$
we can rewrite  $S^{(2)}_{EH_{2}}+S_{\rm int}=S_{B}+S_{W}$ where:
\bees
S_{B}&=&M_{f}^{2}\int d^4x\, \[\mathcal{R}_{\mu\nu}\frac{1}{\Box}\mathcal{R}^{\mu\nu}-\tfrac{1}{2}\mathcal{R}\frac{1}{\Box}\mathcal{R}\] \label{eq:S2a} \\
S_{W}&=&-M_{f}^{2}\int d^4x\, \[ \mathcal{R}_{\mu\nu}\frac{1}{\Box-m^{2}}\mathcal{R}^{\mu\nu}-\tfrac{1}{3}\mathcal{R}\frac{1}{\Box-m^{2}}\mathcal{R}\] \label{eq:S2b}
\ees
The first term can also be rewritten as 
\bees
S_{B}&=&M_{f}^{2}\int d^4x\, \(\mathcal{R}_{\mu\nu}-\tfrac{1}{2}\eta_{\mu\nu}\mathcal{R}\)\frac{1}{\Box}\mathcal{R}^{\mu\nu}\nn\\
&=&M_{f}^{2}\int d^4x\, \mathcal{G}^{\mu\nu}\frac{1}{\Box}\mathcal{R}^{\mu\nu}
\label{eq:Sb}\, ,
\ees
where $\mathcal{G}^{\mu\nu}$ is the linearized Einstein tensor.
As first observed in \cite{Barvinsky:2003kg}, despite its non-local appearance, $S_B$ is local with respect to $\hmn$, and  is just a way of rewriting the quadratic part of the Einstein-Hilbert action. Indeed,  using \eqs{linRmn1}{linRmn2} and performing some integration by parts,
\be
\int d^4x\, \mathcal{G}^{\mu\nu}\frac{1}{\Box}\mathcal{R}^{\mu\nu}=
\frac{1}{4}\int d^4x\,
\hmn \mathcal{E}^{\mu\nu\rho\sigma}h_{\rho\sigma}\, .
\ee
Thus in the end, putting together $S_B+S_W$ with the quadratic Einstein-Hilbert term of the first metric 
$S_{EH_{1}}$, we get
\bees
S_2&\equiv& S^{(2)}_{EH_{1}}+S^{(2)}_{EH_{2}}+S_{\rm int}\\
&=&\frac{M_g^2+M_f^2}{4}\int d^4x\,
\hmn \mathcal{E}^{\mu\nu\rho\sigma}h_{\rho\sigma} 
-M_{f}^{2}\int d^4x\, \[ \mathcal{R}_{\mu\nu}\frac{1}{\Box-m^{2}}\mathcal{R}^{\mu\nu}
-\frac{1}{3}\mathcal{R}\frac{1}{\Box-m^{2}}\mathcal{R}\]\, .\nn
\ees
The non-local term can be rewritten  in terms of the linearized Weyl tensor ${\cal W}^{\mu\nu\rho\sigma}$ observing that
\bees
&&\hspace*{-10mm}
2\( \mathcal{R}_{\mu\nu}\frac{1}{\Box-m^{2}}\mathcal{R}^{\mu\nu}
-\frac{1}{3}\mathcal{R}\frac{1}{\Box-m^{2}}\mathcal{R}\)
={\cal W}_{\mu\nu\rho\sigma} \frac{1}{\Box-m^{2}}{\cal W}^{\mu\nu\rho\sigma}\nn\\
&&- \(  \mathcal{R}_{\mu\nu\rho\sigma}\frac{1}{\Box-m^{2}} \mathcal{R}^{\mu\nu\rho\sigma}
-4 \mathcal{R}_{\mu\nu}\frac{1}{\Box-m^{2}} \mathcal{R}^{\mu\nu}
+ \mathcal{R}\frac{1}{\Box-m^{2}} \mathcal{R} \)\, ,
\ees
Consider now the quantity
\be
\widetilde{\chi}_{E}\equiv \int d^4x\sqrt{-g}\, \(  R_{\mu\nu\rho\sigma}\frac{1}{\Box-m^{2}} R^{\mu\nu\rho\sigma}
-4 R_{\mu\nu}\frac{1}{\Box-m^{2}} R^{\mu\nu}
+ R\frac{1}{\Box-m^{2}} R \)\,.
\ee
If the factor $(\Box-m^{2})^{-1}$ were not present this would be just the Gauss-Bonnet term, which is a topological invariant and does not contribute to the variation of the action. Because of the
$(\Box-m^{2})^{-1}$ factors this is no longer true. However,
expanding over Minkowski space we  find that
\be
\int d^4x\sqrt{-g}\, \[R_{\mu\nu\rho\sigma}\frac{1}{\Box-m^{2}}R^{\mu\nu\rho\sigma}
-4 R_{\mu\nu}\frac{1}{\Box-m^{2}} R^{\mu\nu}
+ R\frac{1}{\Box-m^{2}} R\] = {\cal O}(h^3)\, .
\ee
Therefore in the end, to the order at which we are working, this term can indeed be neglected, and we end up with 
\be\label{S2linWW}
S_2=\frac{M_g^2+M_f^2}{4}\int d^4x\,
\hmn \mathcal{E}^{\mu\nu\rho\sigma}h_{\rho\sigma} 
-\frac{M_{f}^{2}}{2}\int d^4x\,
{\cal W}_{\mu\nu\rho\sigma} \frac{1}{\Box-m^{2}}{\cal W}^{\mu\nu\rho\sigma}\, .
\ee
To the quadratic order at which we are working, this action has the obvious covariantization
\be\label{S2final}
S_2=M^2_{\rm pl}\int d^4x\sqrt{-g}\, R\,
-\frac{M_{f}^{2}}{2}\int d^4x\,\sqrt{-g}\,
W_{\mu\nu\rho\sigma} \frac{1}{\Box-m^{2}}W^{\mu\nu\rho\sigma}\, ,
\ee
where the linearized Weyl tensor ${\cal W}^{\mu\nu\rho\sigma}$ has been promoted to the full Weyl tensor
$W^{\mu\nu\rho\sigma}$, and $M^2_{\rm pl}=M_g^2+M_f^2$.

\section{Discussion}\label{sect:disc}

We conclude with a few comments on our main result, \eq{S2final}. First of all, we observe that, in the limit $m\ra 0$, the result does not reduce to GR. This is a reflection of the vDVZ discontinuity of the original bigravity theory. In fact the original bigravity theory, when linearized over Minkowski, described a massless graviton, plus a massive graviton with a Fierz-Pauli mass term. The bigravity action goes smoothly into the action of GR  in the limit $m\ra 0$, but the discontinuity manifests itself when one computes the propagator. It is quite interesting to observe that, in our non-local formulation, after having eliminated the second metric with its own equations of motion, the discontinuity
manifests itself directly at the level of the action, as we see from \eq{S2linWW}. We  can check that this discontinuity is just the vDVZ discontinuity by computing the propagator associated to the quadratic action 
(\ref{S2linWW}).
Using the explicit expression of the linearized Weyl tensor, \eq{S2linWW} reads
\be\label{S2EF}
S_2=\frac{M_{\rm pl}^2}{4}\int d^4x\, \hmn \[ \mathcal{E}^{\mu\nu\rho\sigma}
-2\tilde{\a}^2 \mathcal{F}^{\mu\nu\rho\sigma}\frac{1}{\Box-m^{2}}\] h_{\rho\sigma} \, ,
\ee
where $\tilde{\a}=M_f/M_{\rm pl}$ and
\bees
\mathcal{F}^{\mu\nu\rho\sigma}&=&\tfrac{1}{3}\partial^{\mu}\partial^{\nu}\partial^{\rho}\partial^{\sigma}-\tfrac{1}{4}\Box(\partial^{\mu}\partial^{\rho}\eta^{\nu\sigma}+\partial^{\mu}\partial^{\sigma}\eta^{\nu\rho}+\partial^{\nu}\partial^{\rho}\eta^{\mu\sigma}+
\partial^{\nu}\partial^{\sigma}\eta^{\mu\rho})\nonumber\\
&&+\tfrac{1}{6}\Box(\partial^{\mu}\partial^{\nu}\eta^{\rho\sigma}+\partial^{\rho}\partial^{\sigma}\eta^{\mu\nu})+\tfrac{1}{4}\Box^2(\eta^{\mu\rho}\eta^{\sigma\nu}+\eta^{\mu\sigma}\eta^{\rho\nu})-\tfrac{1}{6}\Box^2\eta^{\mu\nu}\eta^{\rho\sigma}\, .
\ees
Since the above action is invariant under linearized diffeomorphisms, to invert the quadratic form in \eq{S2EF}
we must add a gauge fixing. Using the usual gauge-fixing term of linearized massless gravity,
${\cal L}_{\rm gf}=-(\paN\bhmn ) (\parho\bar{h}^{\rho\mu})$, where $\bhmn=\hmn -(1/2)h\emn$,
we find, as expected, that the propagator is just the sum of the usual massless propagator of GR plus the propagator 
of a massive graviton with a Fierz-Pauli mass term.
As in  the usual Vainshtein mechanism, the vDVZ discontinuity will then be cured by the non-linearities due to the higher-order terms in the curvature.

Finally, it is interesting to compare \eq{S2final} with the non-local modification of gravity proposed in 
\cite{Maggiore:2014sia}, which is based on the action
\be\label{S1}
S_{\rm NL}=\frac{1}{16\pi G}\int d^{4}x \sqrt{-g}\, 
\[R-\frac{1}{6} m^2R\frac{1}{\Box^2} R\]\, .
\ee
As  discussed in \cite{Maggiore:2014sia,Dirian:2014ara}, this model has quite interesting cosmological properties. Non-local models of this type must be understood as derived from some fundamental non-local theory~\cite{Maggiore:2013mea,Foffa:2013sma},\footnote{The same holds for the non-local model 
proposed in \cite{Deser:2007jk,Deser:2013uya}, see \cite{Woodard:2014iga} for a recent review. This model is however different, since it is rather constructed with a term $Rf(\iBox R)$ in the action, and it does not feature a mass scale $m$. Non-local actions have also been studied with motivation mostly coming from the UV, see e.g.
\cite{Hamber:2005dw,Biswas:2010zk,Biswas:2011ar,Modesto:2011kw,Biswas:2013kla}.
} and it is therefore natural to ask whether they could emerge from bigravity upon elimination of one of the two metrics. We see that the answer is negative. First of all, bigravity produces a different tensor structure, given by the Weyl squared term. Second, as we have seen the non-local term generated from bigravity does not vanish as $m\ra 0$, contrary to the non-local term in \eq{S1}. In retrospect, the fact that the non-local term in \eq{S1} could not have been generated by bigravity is a general consequence of the fact that the theory (\ref{S1}) has no vDVZ discontinuity~\cite{Maggiore:2014sia,Kehagias:2014sda}, while the non-local theory derived from bigravity inherits its vDVZ discontinuity.

\vspace{5mm}
\noindent
{\bf Acknowledgements}.
The work of GC and MM is  supported by the Fonds National Suisse. The work of
JF is supported by an Erasmus grant.

\appendix
\section{Relation to the approach of Hassan, Schmidt-May and von~Strauss}\label{sect:compare}

In this appendix we discuss the relation of our result to that obtained in \cite{Hassan:2013pca}. In general, when we solve the
equations of motion, we can eliminate $\fmn$  using its own equation  of motion, $(\d S/\d\fmn)_g=0$, and then  plugging the resulting expression into $(\d S/\d\gmn)_f=0$ or, alternatively,  we can first obtain $\fmn$ by solving $(\d S/\d\gmn)_f=0$, and  then plug it into $(\d S/\d\fmn)_g=0$. Obviously, these are equivalent and legitimate ways of solving the equations of motion. The issue is more subtle if we want to derive an equivalent effective action involving only $\gmn$. This point has been explained clearly
in~\cite{Hassan:2013pca}: let $S'=S[g,f(g)]$ be the action obtained substituting $\fmn$ with its expression as a function of $\gmn$, obtained either from $(\d S/\d\gmn)_f=0$ or from $(\d S/\d\fmn)_g=0$. In both cases the variation of $S'$ with respect to $\gmn$ is given by 
\be
\frac{\d S'}{\d\gmn (x)}=\(\frac{\d S}{\d\gmn (x)}\)_{f}+\int d^4y\, \frac{\d f_{\rho\sigma}(y)}{\d\gmn (x)}
\(\frac{\d S}{\d f_{\rho\sigma} (y)}\)_{g}=0\, .
\ee
If $\fmn$ is a solution of $(\d S/\d f)_g=0$, then the equations $\d S'/\d g=0$ and $(\d S/\d\gmn)_f=0$ are equivalent. Thus $S'$ is classically equivalent to $S$, as long as we are interested in the dynamics of $\gmn$ only. In contrast, if $\fmn$ is a solution of $(\d S/\d g)_f=0$, the two actions are not equivalent. Solutions of
$(\d S/\d f)_g=0$ do satisfy $\d S'/\d g=0$, but the converse is not necessarily true. The action $S'$ also has spurious solutions 
characterized by $(\d S/\d\fmn)_g=\chi_{\mu\nu}(x)$, with $\chi(x)$ a function such that
\be\label{constraint_chi}
\int d^4y\, \frac{\d f_{\rho\sigma}(y)}{\d\gmn (x)}\chi_{\rho\sigma}(y)=0\, .
\ee
Therefore, in this case extra conditions must be imposed to eliminate the spurious solutions, and the relation between the action $S'$ and the original action is less direct.
On the other hand, the equation $(\d S/\d\gmn)_f=0$ is algebraic in $\fmn$, and can always be solved. In contrast, solving with respect to $\fmn$ the equation $(\d S/\d\fmn)_g=0$ involves the inversion of a differential operator, which in practice can only be done by expanding around a simple background. The approach taken by 
Hassan, Schmidt-May and von~Strauss~\cite{Hassan:2013pca} has been to eliminate  $\fmn$  using the equation of motion of $\gmn$. Plugging the resulting expression into the action $S$ 
it was found in 
ref.~\cite{Hassan:2013pca} that the resulting theory is
given by  the higher-derivative action
\be\label{S2HD}
S_{(2)}^{\rm HD}=M_g^2\int d^4x\sqrt{-g}\,
\[\Lambda +c_R R(g)-\frac{c_{RR}}{m^2}\(\RMN\Rmn-\frac{1}{3}R^2\)\] +{\cal O}(m^{-4})\, ,
\ee
where $\Lambda$, $c_R$ and $c_{RR}$ are some coefficients. By subtracting a Gauss-Bonnet term, similarly to what we have done in sect.~\ref{sect:NLaction},  this action can be rewritten in terms of the Weyl tensor as~\cite{Hassan:2013pca}
\be\label{SHD2}
S_{(2)}^{\rm HD}=M_g^2\int d^4x\sqrt{-g}\,
\[\Lambda +c_R R(g)-\frac{c_{RR}}{2m^2}\, W_{\mu\nu\rho\sigma}W^{\mu\nu\rho\sigma}
\] +{\cal O}(m^{-4})\, .
\ee
The values of $\Lambda$, $c_R$ and $c_{RR}$ are given in \cite{Hassan:2013pca} as functions of the $\beta_i$. With our choice $\beta_0=-6$, $\beta_1=3$, $\beta_2=-1$, $\beta_3=\beta_4=0$ we have\footnote{We also take into account a difference in the definition of $m^2$. Comparing the actions we see that our $m^2$ is related to the parameter denoted $m^2$ in
\cite{Hassan:2013pca} by $m^2_{\rm our}M_f^2=2 m^2_{\rm their} M_g^2$. Observe also that, to determine 
$\Lambda$, $c_R$ and $c_{RR}$, we need to compute the parameter denoted by $a$ in \cite{Hassan:2013pca}, which is determined by their eq.~(2.19). For our values of $\beta_i$ we get $a^2-3a+2=0$, which has the solutions $a=1$ and $a=2$. We only retain $a=1$, since only in this case we get $\Lambda=0$. With $a=2$ not only the coefficient of the Weyl term would differ, but also the cosmological term and the coefficient of the 
Einstein-Hilbert term.}
\be
\Lambda =0\, ,\qquad c_R=\frac{M_g^2+M_f^2}{M^2_g}\, ,
\qquad c_{RR}=4\, \frac{M_g^2+2M_f^2}{M^2_f}\, .
\ee
Comparison with \eq{S2final} shows that the cosmological constant vanishes for both actions, and the coefficient of the Einstein-Hilbert term is the same. However, in the limit $\Box\ll m^2$, the coefficient of the
$W_{\mu\nu\rho\sigma}W^{\mu\nu\rho\sigma}$ term in \eq{S2final} is $+M_f^2/(2m^2)$ while, in \eq{SHD2}, it is $-2 M_g^2 (M_g^2+2M_f^2)/(m^2 M_f^2)$, so the two disagree, even in the sign. Actually, this is simply due to the fact that the action (\ref{SHD2}), by itself, is not equivalent to the original bigravity action, since it has been obtained eliminating $\fmn$ with the equation of motion of $\gmn$, rather than with its own. As discussed in
\cite{Hassan:2013pca}, and has we have recalled above, with this procedure the correspondence between the two theories is more indirect, and is only at the level of the equation of motion, once spurious solutions are eliminated. In contrast, the action (\ref{S2final}) is indeed equivalent to the original bigravity action, up to quadratic orders in the curvature. This point can also be illustrated using a nice example given in app.~A1 of \cite{Hassan:2013pca}. Consider in fact the theory with two scalar fields $\phi$ and $\psi$, given by
\be
S[\phi,\psi]=\int d^4x\, \[-\frac{1}{2}\pam\phi\paM\phi -\frac{1}{2}\pam\psi\paM\psi -\frac{\mu^2}{2}(\phi+\psi)^2\]\, .
\ee
(We do not need source terms for our purpose). Of course, one could diagonalize the action introducing $\Phi_0=\phi-\psi$ and $\Phi_m=\phi+\psi$. However, it is instructive to rather integrate out $\psi$ using either its equation of motion, or the one with respect to $\phi$.
The equation of motions are 
\bees
&&\(\frac{\d S}{\d\phi}\)_{\psi}:\hspace{5mm} (\Box-\mu^2)\phi=\mu^2\psi\, ,\\
&&\(\frac{\d S}{\d\phi}\)_{\phi}:\hspace{5mm} (\Box-\mu^2)\psi=\mu^2\phi\, .
\ees
If we use $(\d S/\d\phi)_{\psi}$, $\psi$ can be eliminated algebraically. Inserting it back in the action one finds the higher-derivative action~\cite{Hassan:2013pca}
\be
S_{\rm HD}=\frac{1}{2\mu^4}\int d^4x\, \phi \Box (\Box-\mu^2)(\Box-2\mu^2)\phi\, ,
\ee
which to first non-trivial order in $\Box/\mu^2$ becomes
\be\label{SHDphi}
S_{\rm HD}\simeq \int d^4x\, \phi\, \Box \(1-\frac{3}{2}\, \frac{\Box}{\mu^2}\)\phi\, .
\ee
If instead we eliminate $\psi$ using its own equation of motion, we have the non-local expression
$\psi=\mu^2 (\Box-\mu^2)^{-1}\phi$. Inserting it in the action we get
\be
S_{\rm non-loc}[\phi]=\frac{1}{2}\int d^4x\,\phi\, \Box\, \(1+\frac{\mu^2}{\mu^2-\Box}\)\phi\, ,
\ee
which,  expanding to first non-trivial order in $\Box/\mu^2$, becomes
\be
S_{\rm non-loc}\simeq \int d^4x\, \phi\,  \Box \(1+\frac{\Box}{2\mu^2}\)\phi\, .
\ee
We see that indeed the first correction is different (even in the sign) from that in \eq{SHDphi}.

\section{Coupling with matter}\label{app:B}

In this appendix we extend the computation of the non-local action to the case of coupling with matter.
In this case \eqs{eq:6a}{eq:6b} become
\bees
\frac{M_{g}^{2}}{M_{f}^{2}}\mathcal{E}_{\mu\nu\rho\sigma}h^{\rho\sigma}+m^{2}(h_{\mu\nu}-\eta_{\mu\nu}h)&=&m^{2}(l_{\mu\nu}-\eta_{\mu\nu}l)-\frac{\kappa_1}{2} T_{\mu\nu}\,,\\
\mathcal{E}_{\mu\nu\rho\sigma}l^{\rho\sigma}+m^{2}(l_{\mu\nu}-\eta_{\mu\nu}l)&=&m^{2}(h_{\mu\nu}-\eta_{\mu\nu}h)-\frac{\kappa_2}{2}  T_{\mu\nu}\,,\, 
\ees
where we have introduced two  generic couplings $\kappa_{1,2} \equiv k_{1,2}/M_f^2$.
Then \eq{eq:10} becomes
\be\label{L}
l_{\mu\nu}=h_{\mu\nu}+\frac{1}{\Box-m^{2}}\[2\mathcal{R}_{\mu\nu}-\frac{1}{3}\(\eta_{\mu\nu}+2\frac{\partial_{\mu}\partial_{\nu}}{m^{2}}\)\mathcal{R}\]-\frac{\kappa_2}{2}\,Q_{\mu\nu\alpha\beta}T^{\alpha\beta}\,,
\ee
where $Q_{\mu\nu\alpha\beta}$ is defined in eq. (\ref{Q}) and
\be\label{l1}
Q_{\mu\nu\alpha\beta}T^{\alpha\beta}=-\frac{1}{\Box-m^2}\left(T_{\mu\nu}-\frac{1}{3}\eta^{\mu\nu}T\right)\,.
\ee
Taking the trace of eq. (\ref{L}) we get
\be\label{l2}
l=h-\frac{2\mathcal{R}}{3m^{2}}-\frac{\kappa_2}{6}\frac{1}{\Box-m^2}T\,.
\ee
We can now substitute back in the action eqs. (\ref{L}) and (\ref{l2}). The result has the form
$S_{\rm tot}=S_{2}+S_{\rm int}$
where $S_2$ is given by eq. (\ref{S2final}) and $S_{\rm int}=\int d^4 x\, \sqrt{-g}\, \mathcal{L}_{\rm int}$ with
\bees
\mathcal{L}_{\rm int}&=&\left(\frac{ k_2}{32 M_f}\right)^2
\left[T_{\mu\nu}\frac{1}{\Box-m^2}T^{\mu\nu}-\frac{1}{3}T\frac{1}{\Box-m^2}T\right]\nn\\
&&-\left(\frac{k_2\, m}{8M_f}\right)^2  \,\left[T_{\mu\nu}\frac{1}{(\Box-m^2)^2}T^{\mu\nu}-\frac{1}{3}T\frac{1}{(\Box-m^2)^2}T\right]\nn\\
&&- k_2 \,m^2\,\left[T_{\mu\nu}\frac{1}{(\Box-m^2)^2}R^{\mu\nu}-\frac{1}{6}T\frac{1}{(\Box-m^2)^2}R\right]\nn\\
&&+\frac{k_2}{2}\left[T_{\mu\nu}\frac{1}{\Box-m^2}R^{\mu\nu}-\frac{1}{6}T\frac{1}{\Box-m^2}R\right]
-\frac{1}{2}(k_1+k_2)\,T^{\mu\nu}\frac{1}{\Box}\,R_{\mu\nu}\,.\label{LintNL}
\ees
The last term can be transformed observing that, at the linearized level, 
\be
\int d^4x\,  T^{\mu\nu}\frac{1}{\Box}\,R_{\mu\nu}=
\int d^4x\, T^{\mu\nu}\frac{1}{\Box}\,(\paR\pam\hnr-\Box\hmn)\, .
\ee
The term $\TMN\iBox\paR\pam\hnr$ vanishes integrating by parts and using $\pam\TMN=0$, and therefore 
in \eq{LintNL} we can replace
$-(1/2)(k_1+k_2)\,T^{\mu\nu}\iBox\,R_{\mu\nu}$ by
$+(1/2)(k_1+k_2)\,T^{\mu\nu}\hmn$, which is the standard local coupling, with an effective Newton constant determined by $k_1+k_2$. The remaining terms in \eq{LintNL} provide genuinely non-local couplings.

\bibliographystyle{utphys}
\bibliography{myrefs_massive}
\end{document}